\documentclass[prd,superscriptaddress,amsfonts,amssymb,amsmath,showpacs,twocolumn]{revtex4-2}
\usepackage{bm}
\usepackage{amsfonts}
\usepackage{latexsym}
\usepackage{graphicx}
\usepackage{amsmath}
\usepackage{palatino}
\usepackage{mathpazo}
\usepackage{textcomp}
\usepackage{float}
\usepackage{booktabs}
\usepackage{dcolumn}
\usepackage{booktabs}
\usepackage{multirow}
\usepackage{hyperref}
\hypersetup{colorlinks,citecolor=blue}
\usepackage{amsmath}
\usepackage{xcolor}
\usepackage{orcidlink}
\usepackage[caption=false]{subfig}
\usepackage{commath}
\usepackage{tabularx}
\def\jnl@style{\it}
\def\aaref@jnl#1{{\jnl@style#1}}

\def\aaref@jnl#1{{\jnl@style#1}}

\def\aj{\aaref@jnl{AJ}}                   
\def\apj{\aaref@jnl{ApJ}}                 
\def\apjl{\aaref@jnl{ApJ}}                
\def\apjs{\aaref@jnl{ApJS}}               
\def\apss{\aaref@jnl{Ap\&SS}}             
\def\aap{\aaref@jnl{A\&A}}                
\def\aapr{\aaref@jnl{A\&A~Rev.}}          
\def\aaps{\aaref@jnl{A\&AS}}              
\def\mnras{\aaref@jnl{Mon.~Not.~Roy.~Astron.~Soc.}}             
\def\prd{\aaref@jnl{Phys.~Rev.~D}}        
\def\prc{\aaref@jnl{Phys.~Rev.~C}}  
\def\prl{\aaref@jnl{Phys.~Rev.~Lett.}}    
\def\qjras{\aaref@jnl{QJRAS}}             
\def\skytel{\aaref@jnl{S\&T}}             
\def\ssr{\aaref@jnl{Space~Sci.~Rev.}}     
\def\zap{\aaref@jnl{ZAp}}                 
\def\nat{\aaref@jnl{Nature}}              
\def\aplett{\aaref@jnl{Astrophys.~Lett.}} 
\def\apspr{\aaref@jnl{Astrophys.~Space~Phys.~Res.}} 
\def\physrep{\aaref@jnl{Phys.~Rep.}}      
\def\physscr{\aaref@jnl{Phys.~Scr}}       
\def\commat{\aaref@jnl{Comm.~Math.~Phys.}}              
\def\science{\aaref@jnl{Science}}               
\def\cqg{\aaref@jnl{Classical Quant.~Grav.}}            
\def\jpcs{\aaref@jnl{JPCS}}                                     
\def\ijmpd{\aaref@jnl{Int.~J.~Mod.~Phys.~D}}                    
\def\grg{\aaref@jnl{Gen.~Relat.~Gravit.}}               
\def\rpp{\aaref@jnl{Rep.~Prog.~Phys.}}          
\def\npa{\aaref@jnl{Nucl.~Phys.~A}}        
\def\lrr{\aaref@jnl{Living Rev.~Rel.}}                   
\def\jcap{\aaref@jnl{J.~Cosmology Astropart.~Phys.}}    
\def\rmp{\aaref@jnl{Rev.~Mod.~Phys.}}   
\def\epjc{\aaref@jnl{Eur.~Phys.~J.~C}}

\allowdisplaybreaks[1]

\begin{document}

\color{black}       

\title{A comprehensive parametrization approach for the Hubble parameter in scalar field dark energy models}

\author{M. Koussour\orcidlink{0000-0002-4188-0572}}
\email[Email: ]{pr.mouhssine@gmail.com}
\affiliation{Quantum Physics and Magnetism Team, LPMC, Faculty of Science Ben
M'sik,\\
Casablanca Hassan II University,
Morocco.} 

\author{N. Myrzakulov\orcidlink{0000-0001-8691-9939}}
\email[Email: ]{nmyrzakulov@gmail.com}
\affiliation{L. N. Gumilyov Eurasian National University, Astana 010008,
Kazakhstan.}
\affiliation{Ratbay Myrzakulov Eurasian International Centre for Theoretical
Physics, Astana 010009, Kazakhstan.}

\author{S. Myrzakulova\orcidlink{0000-0002-0027-0970}}
\email[Email: ]{nmyrzakulov@gmail.com}
\affiliation{L. N. Gumilyov Eurasian National University, Astana 010008,
Kazakhstan.}
\affiliation{Ratbay Myrzakulov Eurasian International Centre for Theoretical
Physics, Astana 010009, Kazakhstan.}

\author{D. Sofuo\u{g}lu\orcidlink{0000-0002-8842-7302}}
\email[Email: ]{degers@istanbul.edu.tr}
\affiliation{Department of Physics, Istanbul University Vezneciler 34134, Fatih, Istanbul, Turkey.}

\date{\today}

\begin{abstract}
This study proposes a novel parametrization approach for the dimensionless Hubble parameter i.e. $E^2(z)=A(z)+\beta (1+\gamma B(z))$ in the context of scalar field dark energy models. The parameterization is characterized by two functions, $A(z)$ and $B(z)$, carefully chosen to capture the behavior of the Hubble parameter at different redshifts. We explore the evolution of cosmological parameters, including the deceleration parameter, density parameter, and equation of state parameter. Observational data from Cosmic Chronometers (CC), Baryonic Acoustic Oscillations (BAO), and the Pantheon+ datasets are analyzed using MCMC methodology to determine model parameters. The results are compared with the standard $\Lambda$CDM model using the Planck observations. Our approach provides a model-independent exploration of dark energy, contributing to a comprehensive understanding of late-time cosmic acceleration.
\end{abstract}

\maketitle

\section{Introduction}
\label{sec1}

In the past two decades, significant progress has been made in observational cosmology, with various observations consistently supporting the notion of an accelerated expansion of the Universe at late times. Type Ia Supernova (SNe) observations \cite{Riess, Perlmutter}, Cosmic Microwave Background (CMB) measurements \cite{R.R., Z.Y.}, and Baryon Acoustic Oscillations (BAOs) studies \cite{D.J., W.J.} have all contributed to reinforcing this understanding. To explain this accelerated expansion, two main approaches have been explored in the literature. The first approach involves introducing an exotic form of matter known as Dark Energy (DE) \cite{Peebles,Padmanabhan,Bambar}. DE is characterized by having a large negative pressure and is postulated to be responsible for driving the observed cosmic acceleration. The second approach considers modifications to the laws of gravity itself \cite{Clifton}. In this context, various modified gravity models have been proposed, such as $f(R)$ gravity \cite{H.A.,H.K., Odintsov1, Odintsov2, Odintsov3,Gu1}, $f(T)$ gravity \cite{T1,T2, Odintsov4, Odintsov5, Gu2}, and $f(Q)$ gravity \cite{Q0,Q1,Q2,Q3,Q4,Gu3}. Furthermore, within the framework of DE models, several candidates have been studied, including quintessence \cite{Quin}, k-essence \cite{ess}, phantom \cite{Phan1,Phan2,Phan3}, and scalar-tensor theories \cite{scal}, each with its own distinct features and implications for the evolution of the Universe. 

While the $\Lambda$CDM model has been remarkably successful in explaining various observational data, it is not without its theoretical challenges \cite{weinberg/1989, CdM1, CdM2}. One of the major issues with the $\Lambda$CDM model is the fine-tuning problem, which arises from the extremely small value of the cosmological constant ($\Lambda$) required to match observations. This fine-tuning problem raises questions about the underlying theoretical framework and why the value of $\Lambda$ is so precisely fine-tuned to produce the observed cosmic acceleration \cite{Copeland}. Another theoretical concern is the cosmic coincidence problem, which refers to the puzzling coincidence that DE density and matter density are comparable at the present epoch, leading to the current accelerated expansion of the Universe. The $\Lambda$CDM model does not provide a natural explanation for this coincidence, and it has been a subject of ongoing debate and investigation \cite{dalal/2001}. Furthermore, recent observations \cite{Sahni0,Delubac,Riess2016,Bonvin} have indicated that the $\Lambda$CDM model may not be the best fit for the most recent low-redshift cosmological data. While the $\Lambda$CDM model remains consistent with a wide range of observations, some data suggests that dynamical DE models, where the DE density evolves with time, might provide a more accurate description of the observed Universe. These theoretical and observational motivations have led researchers to explore dynamical DE models that can address the limitations of the $\Lambda$CDM model. 

Indeed, while various dynamical DE and modified gravity models have been proposed to explain late-time cosmic acceleration, it remains essential to rigorously analyze and scrutinize these models using cosmological observations. The need for a detailed analysis arises from the complexity of the underlying physics governing late-time acceleration and the ever-increasing precision of observational data. To perform such analyses, it is crucial to adopt appropriate parametrizations that allow for model-independent descriptions of the late-time cosmic acceleration \cite{Shafieloo1,Shafieloo2,Dinda,Corasaniti}. Model-independent parametrizations are particularly valuable because they do not assume a specific theoretical framework, enabling a more agnostic approach to understanding the observed data. Recently, Roy et al. \cite{Roy} investigated the nature of DE using a novel parametrization of the Hubble parameter. The authors explored scalar field DE models, including quintessence and phantom, through a model-independent approach. Pacif et al. \cite{Pacif} presented an accelerating cosmological model based on a parametrization of the Hubble parameter. The authors introduced a novel approach to describe the expansion rate of the Universe, leading to an accelerating phase. In their study, Koussour et al. \cite{Koussour1} proposed a novel parametrization of the Hubble parameter i.e. $H^2\left( z\right) =H_{0}^2\left[\left( 1-\alpha \right) +\left( 1+z\right)
\left( \alpha +\beta z\right)\right]$ within the framework of $f(Q)$ gravity. The authors explored the behavior of the Hubble parameter using this new parametrization, shedding light on the late-time cosmic acceleration and its implications in $f(Q)$ gravity models. Sahni et al. introduced the concept of the statefinder, a novel geometric tool for diagnosing the nature and behavior of DE, utilizing a model-independent parameterization of $H(z)$, i.e. $H^2\left( z\right) =H_{0}^2\left[\Omega_{m0}(1+z)^3+A+B(1+z)+C(1+z)^2 \right]$  \cite{Sahni}. Cunha and Lima conducted an investigation into the concept of the transition redshift, which signifies the shift from deceleration to acceleration in cosmic expansion. They presented novel kinematic constraints derived from supernovae observations, employing two different parameterizations of the deceleration parameter: $q(z)=q_{0}+q_{1}z$ and $q(z)=q_{0}+q_{1}z(1 +z)-1$. Their analysis revealed a transition redshift value of $z_{tr}=0.61$ \cite{Cunha1,Cunha2}. Mamon conducted an investigation into the reconstruction of the interaction rate within the holographic DE model. This study employed the Hubble horizon as the infrared cut-off and focused on a specific parameterization of the effective EoS parameter: $\omega_{eff}(z)=-1+\frac{A}{A+B (1+z)^{-n}}$ \cite{Mamon1}. In the same context, numerous parametrizations have been proposed for various physical and geometrical parameters \cite{Jassal,Barboza,Chevallier,Campo,Log1}.

The issue with most of these parametrizations lies in their behavior at extreme values of redshift. In the far future, many parametrizations lead to divergent results for the deceleration parameter, which can cause inconsistencies in predicting the long-term evolution of the Universe. On the other hand, some of these parametrizations are restricted to low redshifts (i.e., $z<<1$) and fail to provide accurate predictions for earlier cosmic epochs \cite{Gong}. Motivated by the aforementioned discussions and the need for a robust and versatile parametrization approach, this study focuses on exploring scalar field DE models using a general scheme based on the dimensionless Hubble parameter. The unique aspect of this approach is that it allows us to express the relevant cosmological parameters in a form that is independent of the specific nature of the scalar field. By adopting this comprehensive parametrization approach for the dimensionless Hubble parameter i.e. $E^2(z)=\frac{H^2(z)}{H_{0}^2}=A(z)+\beta (1+\gamma B(z))$ (where $\beta$, $\gamma$ are free parameters, and $A(z)$, $B(z)$ are functions of the redshift $z$), we aim to address the challenges associated with existing parametrizations that diverge in the far future or are limited to low redshifts. The dimensionless Hubble parameter plays a central role in understanding the Universe's evolution, making it a crucial and significant parameter to investigate in cosmology. The functions $A(z)$ and $B(z)$ are chosen in a manner that allows the parametrization to capture the behavior of the Hubble parameter at various redshifts. For this purpose, we use measurements from Cosmic Chronometers (CC), BAOs, and Pantheon+, which includes an expanded dataset of Pantheon from Type Ia SNe. This dataset comprises 1701 light curves from 1550 Type Ia SNe, collected from different studies. The numerical analysis is conducted using MCMC methods. Furthermore, we compare our parametrization with the $\Lambda$CDM model using Planck observations.

This paper is structured as follows: Sec. \ref{sec2} presents the fundamental mathematical formulation and the dynamics of the scalar field. It introduces a comprehensive parametrization approach for the dimensionless Hubble parameter $E(z)$. Furthermore, this section derives analytical expressions for various cosmological parameters associated with this parametrization. Sec. \ref{sec3} of this paper focuses on the observational data obtained from diverse sources, including Cosmic Chronometers, BAO, and the recently released Pantheon+ datasets. The methodology employed for determining the model parameters is also discussed in this section. The summary of the evolution of the cosmological parameters is provided in Sec. \ref{sec4}. Finally, in Sec. \ref{sec5}, we present a concise summary and draw our conclusions based on the findings from this study.

\section{Cosmological model}
\label{sec2}

For a spatially flat, homogeneous, and isotropic Friedmann-Lemaître-Robertson-Walker (FLRW) Universe described by the metric \cite{Ryden}
\begin{equation}
ds^{2}=dt^{2}-a^{2}(t)[dr^{2}+r^{2}(d{\theta }^{2}+sin^{2}\theta d{\phi }%
^{2})],  \label{FLRW}
\end{equation}%
where $a(t)$ is the scale factor of the Universe as a function of time, we consider the presence of two perfect fluids. The first fluid represents ordinary matter, with negligible pressure, and the second fluid is a scalar field, which is considered as a candidate for DE. In this scenario, Einstein's field equations and the Klein-Gordon equation for the scalar field can be written as follows (assuming $8\pi G=c=1$)
\begin{eqnarray}
  3H^2=\rho_{m} +\rho_{\phi}=\rho_{m}+\frac{1}{2} \dot{\phi^2}+V(\phi),\label{F1}\\
  2\dot{H} +3H^2= -p_{\phi}= -\frac{1}{2} \dot{\phi^2} + V(\phi),\label{F2}\\
   \Ddot{\phi}+3 H\dot{\phi}+  \frac{dV}{d\phi} = 0. \label{KG}
\end{eqnarray} 

Here, $H=\frac{\dot{a}}{a}$ represents the Hubble parameter, which characterizes the rate of cosmic expansion. The symbols $\rho_{m}$, $\rho_{\phi}$, and $p_{\phi}$ denote the energy densities of ordinary matter, the scalar field (DE), and the pressure associated with the scalar field, respectively. By solving these equations and imposing suitable initial conditions, one can investigate the cosmic expansion history and the behavior of the scalar field over cosmic time. 

Further, the energy density for ordinary matter evolves with the scale factor as 
\begin{equation}
    \rho_{m}=\rho_{m0}a^{-3}=\rho_{m0}(1+z)^{3},
\end{equation}
where $\rho_{m0}$ represents the current value of matter-energy density. In addition, $z$ denotes the redshift parameter, defined as $z=\frac{1}{a(t)}-1$. On the other hand, the energy density for the scalar field can be expressed a $\rho_{\phi}=\frac{1}{2} \dot{\phi^2} +V(\phi)$, and the corresponding pressure component as $p_{\phi}=\frac{1}{2} \dot{\phi^2} - V(\phi)$ \cite{Barrow1,Barrow2}. The function $V(\phi)$ represents the potential associated with the scalar field $\phi$.

By manipulating Eqs. (\ref{F1}), (\ref{F2}), and (\ref{KG}), one can derive expressions for the derivative of the Hubble parameter $\dot{H}$ and the scalar field potential $V(\phi)$ as
\begin{equation}\label{Hdot}
2\dot{H} = - \frac{\rho_{m0}}{a^3}- \dot{\phi^2 },
\end{equation}
and
\begin{equation}\label{Vphi}
V(\phi) = \dot{H}+3 H^2-\frac{\rho_{m0}}{2 a^3}. 
\end{equation}

By performing further manipulations of Eq. (\ref{Hdot}) and employing the standard relation $\dot{H}=\frac{1}{2}a\frac{d}{da}(H^2)$, we obtain: $a\frac{d}{da}(H^2) + \frac{\rho_m{}_0}{a^3} = -\epsilon\dot{\phi^2}$. In addition, we can express $\dot{\phi}$ using the previous equations as $\dot{\phi}=a H\left(\frac{d\phi}{da}\right)$. Now, we can find the derivative of the scalar field $\phi$ with respect to redshift $z$ as
\begin{equation}\label{dphidzE}
  \frac{d\phi}{dz} =\left[\frac{2E\frac{dE}{dz} -3\Omega_m{}_0(1+z)^2}{ E^2(1+z)}\right]^\frac{1}{2}. 
\end{equation}

Here, $E \equiv E(z)$ represents the dimensionless Hubble parameter ($E(z)=\frac{H(z)}{H_{0}}$), where $H_{0}$ is the present value of Hubble parameter, and $\Omega_m{}_0=\frac{\rho_m{}_0}{3H_{0}^2}$ denotes the present-day density parameter of matter.

Likewise, applying a similar approach, we can express the scalar field potential, as given in Eq.  (\ref{Vphi}) in terms of $z$ as
\begin{equation}\label{VphiE}
\frac{V(z)}{3H_{0}^2}=-\frac{(1+z)}{3}E\frac{dE}{dz}+E^2-\frac{1}{2}\Omega_{m0}(1+z)^3.
\end{equation}

To characterize the nature of the cosmological expansion, whether it is accelerating or decelerating, we introduce the deceleration parameter $q$, which is defined as
\begin{equation}\label{q}
q(z)=-1-\frac{\dot{H}}{H^2}=\frac{(1+z)}{E}\frac{dE}{dz}-1.
\end{equation}

When $q>0$, this signifies a deceleration in the expansion of the Universe. For $q=0$, the expansion maintains a constant rate. Conversely, when $-1 <q < 0$, it indicates accelerating growth. Notably, when $q = -1$, the Universe experiences exponential expansion, known as de Sitter expansion. Moreover, for $q < -1$, the Universe exhibits super-exponential expansion.

The density parameter for the matter field $\Omega_{m}$, and the density parameter for the scalar field $\Omega_{\phi}$, are crucial cosmological parameters that offer significant information about the matter composition of the Universe,
\begin{equation}\label{Omegam}
 \Omega_m(z)=\frac{\rho_{m}}{3H^2}=\frac{\Omega_{m0}(1+z)^3}{E^2},
\end{equation}
\begin{equation}\label{Omegaphi}
\Omega_{\phi}(z)=1-\Omega_{m}(z)=1-\frac{\Omega_{m0}(1+z)^3}{E^2},
\end{equation}

To gain a deeper understanding of the accelerated period, we introduce the equation of state (EoS) parameter $\omega_{\phi}(z)$, which is defined as
\begin{equation}
\omega_{\phi}(z)=\frac{p_{\phi}}{\rho_{\phi} }=\frac{-1-\frac{2 \dot{H}}{3 H^2}}{\Omega_{\phi}}.
\end{equation}

Consequently, this leads to
\begin{equation}\label{wphi}
\omega_{\phi}(z)=\frac{\frac{2}{3}(1+z)E\frac{dE}{dz}-E^2}{E^2-\Omega_{m0}(1+z)^3},
\end{equation}

From Eqs. (\ref{F1}) and (\ref{F2}), we can derive the equation that describes the acceleration, as mentioned in \cite{Sahni1},
\begin{equation}
    \frac{\overset{..}{a}}{a}=-\frac{1}{6}\left( \rho_{m} +\rho_{\phi}+3p_{\phi}\right).
\end{equation}

Therefore, based on the derived expression, the current model predicts acceleration ($\overset{..}{a}>0$) only when $\omega<-\frac{1}{3}$. During this accelerated phase of evolution, three distinct periods can be identified, characterized by the value of  $\omega$ \cite{Myrzakulov}:  
\begin{itemize}
    \item $-1/3 < \omega< -1$: In this phase, DE behaves like a quintessence field, with the pressure being greater than $-\frac{1}{3}$ but less than $-1$. The scalar field associated with DE evolves gradually, leading to cosmic acceleration.  
    \item $\omega = -1$: This value represents the cosmological constant in the $\Lambda$CDM model, associated with DE. It signifies a constant negative pressure that remains fixed over time, driving the cosmic acceleration in the $\Lambda$CDM model.
    \item $\omega< -1$: This signifies the onset of the phantom era, indicating DE with even stronger negative pressure than the cosmological constant. The phantom era represents exotic DE characterized by a rapidly evolving scalar field, resulting in the accelerated expansion of the Universe.
\end{itemize}

Further, the parametrization of the dimensionless Hubble parameter $E(z)$ plays a crucial role in characterizing the nature of the Universe's expansion rate. In a general setting, $E(z)$ can be expressed as
\begin{equation}
\label{Ez}
    E^2(z)=A(z)+\beta (1+\gamma B(z)),
\end{equation}
where $\beta$, $\gamma$ are free parameters, and $A(z)$, $B(z)$ are functions of the redshift $z$. Indeed, the literature has witnessed the proposal of various functional forms for $A(z)$ and $B(z)$, aiming to address cosmological problems effectively \cite{Cunha1,Cunha2,Chevallier,Jassal,Barboza,Mamon1,Campo,Log1}. However, as previously mentioned, some of these parameterizations suffer from a lack of predictive capability concerning the future evolution of the Universe, while others are valid only for low redshift. In Ref. \cite{Log2}, the authors introduced a parametrization of
$E(z)$ that includes two correction terms associated with DE in the context of the $\Lambda$CDM model. Their goal was to study the entire expansion history of the Universe. Notably, they demonstrated that this model aligns better with current observational constraints when certain restrictions on model parameters are imposed. Despite these efforts, the search for an appropriate functional form of $E(z)$ that can adeptly address cosmological challenges remains ongoing. Researchers continue to explore novel parametrizations that can account for the complexities of the Universe's expansion and offer a comprehensive understanding of its dynamics.

In our study, the chosen dimensionless parametrization in Eq. (\ref{Ez}) is favored for its flexibility in characterizing the Hubble parameter's behavior across various redshifts. The choice of parametrization is motivated by the need to understand the late-time cosmic acceleration and the nature of DE without being tied to a specific theoretical framework. The functions $A(z)$ and $B(z)$ are chosen in a manner that allows the parametrization to capture the behavior of the Hubble parameter at various redshifts. By adjusting the value of the free parameters $\beta$, $\gamma$, and the functional forms of $A(z)$ and $B(z)$, this parametrization can accommodate different cosmological models and allow for comparisons with observational data. To obtain the $\Lambda$CDM model using this parametrization, we need to set the appropriate values for $\beta$, $\gamma$, and $A(z)$. The $\Lambda$CDM model is a specific case of this parametrization, and the values are chosen as follows: $\beta=\Omega_{\Lambda}$, $\gamma=0$, $A(z)=\Omega_{m0}(1+z)^3$. Based on these considerations, in this current study, we introduce a novel parametrization of the dimensionless Hubble parameter. This parametrization includes correction terms associated with DE in the context of the $\Lambda$CDM model, and it can be expressed as $A(z)=\alpha(1+z)^3$ and $B(z)=\frac{z}{1+z}$, where $\alpha=\Omega_{m0}$ when $\gamma=0$. Since $B(z=0)=0$, it imposes an additional constraint on the parameters of the model, reducing their number and leading to the relationship $\alpha+\beta=1$.

Considering the specific choice of $E(z)$ as given in Eq. (\ref{Ez}), Eqs. (\ref{q}), (\ref{Omegam}), (\ref{Omegaphi}), and (\ref{wphi}) can be expressed as 
\begin{equation}\label{q1}
q(z)=\frac{3 \alpha  (1+z)^3+(1-\alpha )\frac{\gamma}{(1+z)}}{\left[\alpha  (1+z)^3+(1-\alpha ) \left(1+\frac{\gamma  z}{1+z}\right)\right]^{1/2}}-1,
 \end{equation}
\begin{equation} \label{omegam1}
\Omega_{m}(z)=\frac{\Omega_{m0} (1+z)^3}{\alpha  (1+z)^3+(1-\alpha ) \left(1+\frac{\gamma  z}{1+z}\right)},
\end{equation}
   \begin{equation}
  \Omega_{\phi}(z)=1-\frac{\Omega_{m0} (1+z)^3}{\alpha  (1+z)^3+(1-\alpha ) \left(1+\frac{\gamma  z}{1+z}\right)},
\end{equation}
and
  \begin{equation} \label{wphi1}
 \omega_{\phi}(z) =\frac{\alpha \left( 1+z\right) ^{2}\left( \alpha \left( 1+z\right) -1\right)
+\left( 1-\alpha \right) \left( 1+\frac{\gamma \left( 1+3z\right) }{3\left(
1+z\right) }\right) }{\left[ \alpha
\left( 1+z\right) ^{3}+\left( 1-\alpha \right) \left( 1+\frac{\gamma z}{1+z}%
\right) \right]-\Omega_{m0}(1+z)^3}.
\end{equation}

To ensure comprehensiveness, by substituting Eq. (\ref{Ez}) into Eqs. (\ref{dphidzE}) and (\ref{VphiE}), we have derived the expressions for the potential $\phi(z)$ and $V(\phi)$ specific to this particular selection of $E(z)$,
\begin{equation}
\phi (z)=\int{\left[
\frac{3\alpha \left( 1+z\right) +\left( 1-\alpha \right) \frac{\gamma }{%
\left( 1+z\right) ^{3}}-6\Omega _{m0}\left( 1+z\right) }{2\left[ \alpha
\left( 1+z\right) ^{3}+\left( 1-\alpha \right) \left( 1+\frac{\gamma z}{1+z}%
\right) \right] }
\right]^\frac{1}{2}} dz,
\end{equation}
and
\begin{equation}\label{eq:vphi}
\begin{split}
\frac{V(z)}{3H_{0}^2}=-\frac{3\alpha \left( 1+z\right) ^{2}+\left( 1-\alpha \right) \frac{\gamma }{%
1+z}}{6}+\\
\left[\alpha
\left( 1+z\right) ^{3}+\left( 1-\alpha \right) \left( 1+\frac{\gamma z}{1+z}%
\right)\right]-\frac{1}{2}\Omega_{m0}(1+z)^3.
\end{split}
\end{equation}

In the next section, a statistical analysis was conducted to constrain the parameters ($H_0,\alpha,\gamma$) of the model. Using the best-fit values obtained for these parameters, the evolution of various relevant cosmological parameters was thoroughly investigated.

\section{Analysis of Observational Data and Methodology}
\label{sec3}

Observational cosmology is characterized by the crucial task of constructing optimal cosmological models. To achieve this, it is essential to rigorously constrain the model parameters, namely $\alpha$, $\gamma$, and the present value of the Hubble parameter $H_{0}$, through meticulous analysis of observational data. In this study, we use a diverse array of observational datasets, which include CC, BAO, and the latest Pantheon sample known as Pantheon+, obtained from observations of SNe.

\subsection{$Hz$ dataset: CC}
The CC method constitutes a valuable technique utilized to determine the Hubble rate by studying the characteristics of the most ancient and passively evolving galaxies. These galaxies are meticulously chosen based on a narrow redshift interval, enabling the application of the differential aging method. The Hubble rate $H(z)$ defined within the FLRW metric, is given by the expression: $H = -\frac{1}{1+z} \frac{dz}{dt}$. The provided relationship enables us to deduce the rate of the Universe's expansion at various time points. A significant advantage of the CC method lies in its capability to measure the Hubble parameter $H(z)$ without being dependent on specific cosmological assumptions. This feature makes the CC method a valuable tool for testing and critically examining different cosmological models. Notably, Jimenez and Loeb \cite{jimlo} introduced a procedure that directly obtains Hubble parameter data by calculating the rate of redshift change $dz/dt$, at a precise value of $z$. This direct approach enhances the precision of the Hubble parameter measurements and contributes to a more robust analysis of the Universe's expansion history.

For this investigation, a meticulous compilation of a comprehensive dataset comprising 31 data points has been undertaken from a variety of reputable surveys (readers may refer to Table 3 in \cite{DMNaik}). These data points are obtained using the CC method and encompass a wide range of redshift values, spanning from 0.1 to 2. The subsequent analysis employs the Markov Chain Monte Carlo (MCMC) technique, incorporating the use of the $\chi^2$ function to analyze cosmic chronometers. The $\chi^2$ function is expressed as:
\begin{equation}
\chi^2_{H(z)}(H_{0},\alpha,\gamma)=\sum_{k=1}^{31}\left[\frac{(H_{th}(z_k,H_{0},\alpha,\gamma)-H_{obs}(z_k))^2)}{\sigma_H^2(z_k)}\right].
\end{equation}

Here, $H_{th}$ represents the theoretical estimation of the Hubble parameter for a specific model, characterized by model parameters $H_{0}$, $\alpha$, and $\gamma$. $H_{obs}$ denotes the observed values of the Hubble parameter, and $\sigma_H$ represents the associated error in the estimation.

\subsection{$BAO$ dataset}

BAOs manifest as fluctuations in the density of baryonic matter throughout the Universe, originating from acoustic density waves in the primordial plasma during its early stages. These oscillations offer valuable insights as they can be harnessed to extract significant cosmological parameters related to DE. In this study, we include the dataset from BAOs, which has been collected from various surveys, including the 6dFGS, the SDSS, and the LOWZ samples of the BOSS \cite{BAO1,BAO2,BAO3,BAO4,BAO5,BAO6}. These surveys have yielded exceptionally accurate measurements of the positions of BAO peaks in galaxy clustering across different redshifts. The characteristic scale of BAO, represented by the sound horizon $r_s$ at the epoch of photon decoupling with redshift $z_{dec}$, is related through the following equation:
\begin{equation}
r_{s}(z_{\ast })=\frac{c}{\sqrt{3}}\int_{0}^{\frac{1}{1+z_{\ast }}}\frac{da}{
a^{2}H(a)\sqrt{1+(3\Omega _{b,0}/4\Omega _{\gamma,0})a}},
\end{equation}
where $\Omega _{b,0}$ and $\Omega _{\gamma,0}$ represent the present density values of baryons and photons, respectively. The BAO dataset used in this study comprises six data points for the ratio $d_{A}(z_{\ast })/D_{V}(z_{BAO})$. These data points were obtained from the sources cited in Refs. \cite{BAO1, BAO2, BAO3, BAO4, BAO5, BAO6}. Here, $z_{\ast }\approx 1091$ represents the redshift value for photon decoupling, and $d_{A}(z_{\ast })=c\int_{0}^{z}\frac{dz'}{H(z')}$ represents the comoving angular diameter distance at decoupling. In addition, we have the dilation scale $D_{V}(z)=\left[czd_{A}^{2}(z)/H(z)\right] ^{1/3}$. 

The chi-square function, introduced in \cite{BAO6}, is used to evaluate the BAO dataset, and it can be expressed as
\begin{equation}\label{4e}
\chi _{BAO}^{2}=X^{T}C_{BAO}^{-1}X,
\end{equation}
where 
\begin{equation}
X=\left( 
\begin{array}{c}
\frac{d_{A}(z_{\star })}{D_{V}(0.106)}-30.95 \\ 
\frac{d_{A}(z_{\star })}{D_{V}(0.2)}-17.55 \\ 
\frac{d_{A}(z_{\star })}{D_{V}(0.35)}-10.11 \\ 
\frac{d_{A}(z_{\star })}{D_{V}(0.44)}-8.44 \\ 
\frac{d_{A}(z_{\star })}{D_{V}(0.6)}-6.69 \\ 
\frac{d_{A}(z_{\star })}{D_{V}(0.73)}-5.45%
\end{array}%
\right) \,,
\end{equation}%
and $C_{BAO}^{-1}$ represents the inverse of the covariance matrix \cite{BAO6}.

\subsection{\textit{SNe} dataset: Pantheon+}

The Pantheon+ analysis goes beyond the original Pantheon analysis by incorporating an extended dataset of SNe that includes those with measured Cepheid distances to galaxies. This comprehensive dataset consists of 1701 light curves from 1550 SNe, covering a redshift range of $0.001\le z\le 2.2613$, and has been collected from a total of 18 distinct studies \cite{pan1,pan2,pan3,pan4}. 
Among the 1701 light curves present in the dataset, 77 of them are linked to galaxies that contain Cepheids. The Pantheon+ compilation, in comparison to the original Pantheon compilation by \cite{Scolnic}, brings substantial enhancements and improvements. Mainly, the Pantheon+ compilation showcases an enlarged sample size, with a notable increase in the number of SNe at redshifts below 0.01. 
Moreover, substantial enhancements have been made to address and mitigate systematic uncertainties related to redshifts, intrinsic scatter models, photometric calibration, and peculiar velocities of SNe. It is essential to note that, due to specific selection criteria, not all SNe from the original Pantheon compilation are included in the improved Pantheon+ compilation.

Pantheon+ presents an additional advantage by enabling the constraint of the Hubble constant ($H_0$) alongside model parameters. To obtain the best fits for the free parameters, the optimization of the $\chi^2$ function is necessary, as expressed below:
\begin{equation}
    \chi^2_{SNe}= \Delta\mu^T (C_{Sys+Stat}^{-1})\Delta\mu.
\end{equation}

In this context, $C_{Sys+Stat}$ corresponds to the covariance matrix of the Pantheon+ dataset, encompassing both systematic and statistical uncertainties. 

The term $\Delta\mu$ signifies the distance residual and is defined as follows:
\begin{equation}
    \Delta\mu_k=\mu_k-\mu_{th}(z_k).
\end{equation}

Here, $\mu_k$represents the distance modulus of the $k^{th}$ SNe. It is crucial to note that $\mu_k$ is calculated as $\mu_k=m_{Bk}-M$, where $m_{Bk}$ corresponds to the apparent magnitude of the $k^{th}$ SNe and $M$ denotes the fiducial magnitude of a SNe.

The theoretical distance modulus $\mu_{th}$ is calculated using the following expression:
\begin{equation}\label{Eq:mu}
    \mu^{th}(z,H_{0},\alpha,\gamma)=5\log_{10}\left(\frac{d_L(z,H_{0},\alpha,\gamma)}{1\text{ Mpc}} \right)+25,
\end{equation}
where $d_L$ represents the luminosity distance in Mpc, which is model-based and given by:
\begin{equation}
d_L(z,H_{0},\alpha,\gamma)=\frac{c(1+z)}{H_0}\int_0^z \frac{dy}{E(y)},
\end{equation}
where $c$ represents the speed of light. Furthermore, the parameters $M$ and $H_0$ display degeneracy, particularly in the analysis of SNe, but considering the recent SH0ES results relaxes these constraints. As a result, the distance residual can be expressed as:
\begin{equation}
    \Delta\Bar{\mu}=\begin{cases}
            \mu_k-\mu_k^{cd}, & \text{if $k$ is in Cepheid hosts}\\
            \mu_k-\mu_{th}(z_k), & \text{otherwise}
         \end{cases},
\end{equation}
where $\mu_k^{cd}$ represents the Cepheid host-galaxy distance released by SH0ES. When calculating the covariance matrix for the Cepheid host-galaxy, it can be combined with the covariance matrix for SNe. The resulting combined covariance matrix, denoted as $C^{SNe}_{Sys+Stat}+C^{cd}_{Sys+Stat}$, ncorporates both statistical and systematic uncertainties from the Pantheon+ dataset and Cepheid host-galaxy data.  Thus, the $\chi^2$ function for the combined covariance matrix used to constrain cosmological models in the analysis is given by
\begin{equation}\label{eq:chiSNe}
    \chi^2_{SNe+}= \Delta\Bar{\mu} (C^{SNe}_{Sys+Stat}+C^{cd}_{Sys+Stat})^{-1}\Delta\Bar{\mu}^T.
\end{equation}

\subsection{Joint Analysis}
Finally, we explore various combinations of the aforementioned observational datasets. The following combinations will be employed for our study:
\begin{eqnarray}
 Hz+BAO,\\
 Hz+BAO+SNe.
\end{eqnarray} 

\begin{figure}[!]
     \centering
     \includegraphics[width=\linewidth]{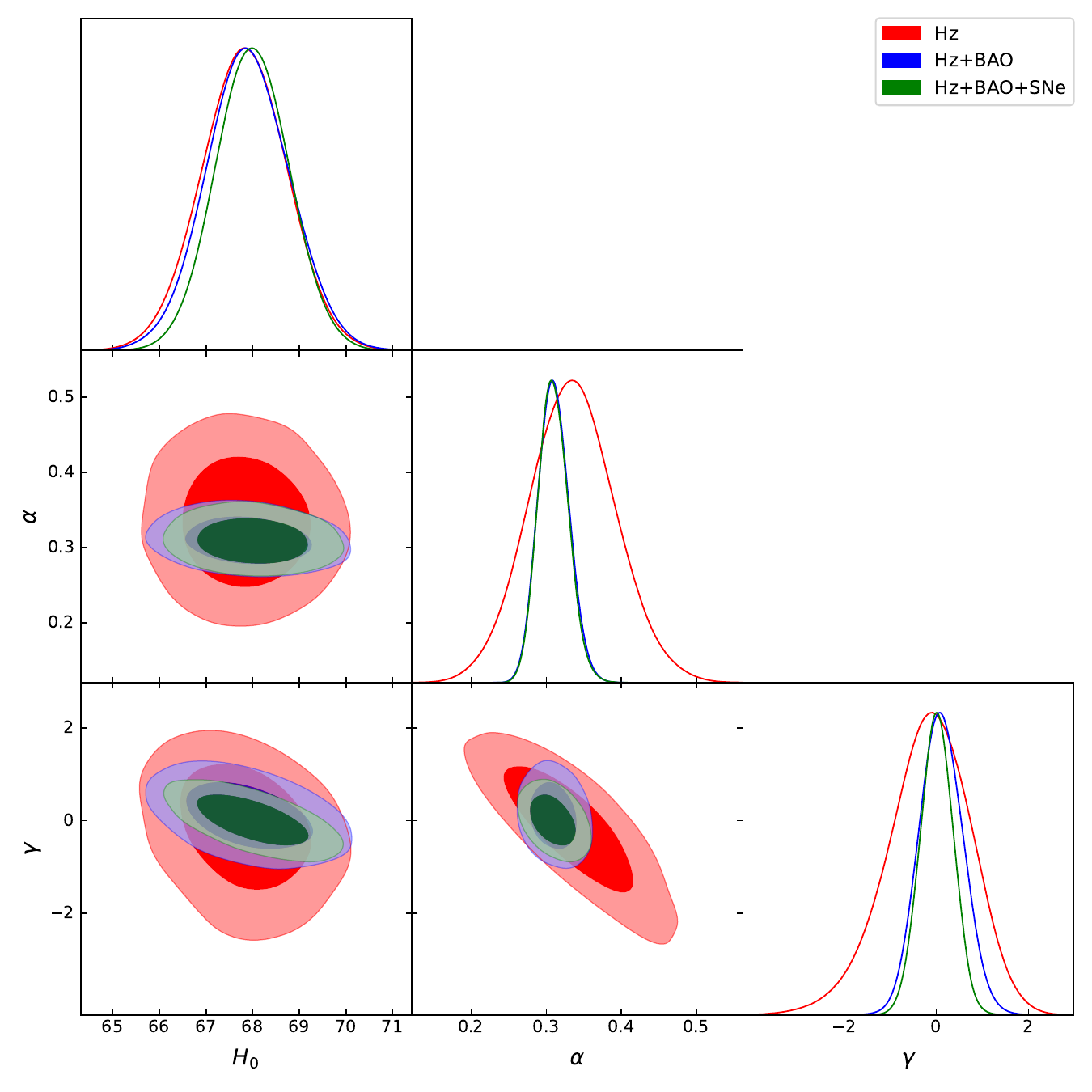}
     \caption{Likelihood contour for $Hz$, $Hz+BAO$, and $Hz+BAO+SNe$ datasets with $1-\sigma$ and $2-\sigma$ confidence levels.}\label{combine}
\end{figure}

The model parameters are constrained by minimizing their respective $\chi^2$ values, which are related to the likelihood through $\mathcal{L} \propto \exp \left( -\frac{\chi^2}{2} \right)$, using the Markov Chain Monte Carlo (MCMC) sampling technique and the {\tt emcee} library. The obtained results are summarized in Tab. \ref{tab1}. Fig. \ref{combine} depicts the $1-\sigma$ and $2-\sigma$ contour plots for $Hz$, $Hz+BAO$, and $Hz+BAO+SNe$, respectively.

\begin{table*}[t]
    \caption{Summary of MCMC analysis results for parameters $H_0$, $\alpha$, and $\gamma$.}
    \label{tab1}
	\centering
	\begin{tabularx}{\linewidth}{>{\centering\arraybackslash}X >{\centering\arraybackslash}X >{\centering\arraybackslash}X >{\centering\arraybackslash}X}
        \hline
        \hline
		$Dataset$ &  $H_0$ & $\alpha=\Omega_{m0}$ & $\gamma$\\ 
		\hline
            $Hz$ & $67.8\pm 1.8$ & $0.33\pm0.11$ & $-0.2\pm1.8$\\
            $Hz+BAO$ & $67.9_{-1.7}^{+1.8}$ & $0.310_{-0.038}^{+0.042}$ & $0.10_{-0.90}^{+0.94}$ \\
            $Hz+BAO+SNe$ & $68.0_{-1.5}^{+1.6}$ & $0.309_{-0.038}^{+0.041}$ & $0.01_{-0.71}^{+0.70}$ \\
            \hline
            \hline
	\end{tabularx}
    \end{table*}

\section{Results from Data Analysis}
\label{sec4}
\begin{figure}[h]
\includegraphics[scale=0.7]{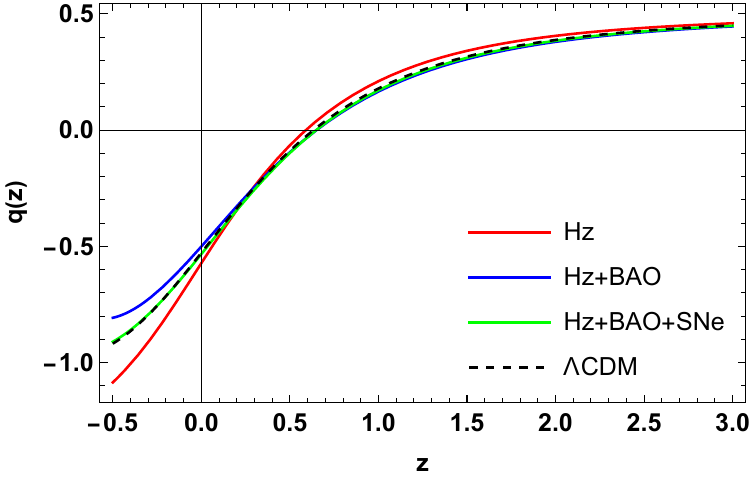}
\caption{This figure shows the evolution the deceleration parameter $q(z)$ against $z$ for constrained values from $Hz$, $Hz+BAO$, and $Hz+BAO+SNe$ datasets. In additon, we include a comparison to the $\Lambda$CDM model using the parameter values from Planck observations \cite{Planck2020}.}\label{F_q}
\end{figure}

In the preceding section, we analyzed the observational constraints on the parameters $H_0$, $\alpha$, and $\gamma$. Now, let us explore the evolution of the cosmological parameters based on these constraints. In our analysis, we specifically concentrate on three datasets: $Hz$, $Hz+BAO$, and $Hz+BAO+SNe$. Using the best-fit values of $H_0$, $\alpha$, and $\gamma$ (Tab. \ref{tab1}), we reconstruct the deceleration parameter $q(z)$, and the results are presented in Fig. \ref{F_q}. The plot clearly shows that $q(z)$ indicates past deceleration ($q > 0$) and recent acceleration ($q < 0$) of the Universe. This is crucial for understanding the structure formation of the Universe. The present values of $q(z)$ are found to be $q_0=-0.57_{-0.46}^{+0.46}$, $q_0=-0.50_{-0.35}^{+0.36}$, and $q_0=-0.53_{-0.29}^{+0.29}$ for $Hz$, $Hz+BAO$, and $Hz+BAO+SNe$ datasets, respectively. In addition, the best-fit values for the transition redshift ($z_{tr}$) are determined as $z_{tr}=0.59_{-0.4}^{+0.4}$, $z_{tr}=0.65_{-0.06}^{+0.08}$, and $z_{tr}=0.65_{-0.08}^{+0.09}$ for $Hz$, $Hz+BAO$, and $Hz+BAO+SNe$ datasets, respectively. These results are consistent with previous findings by various researchers using different approaches \cite{Nair,Wang,Koussour2,Hernandez}. In Fig. \ref{F_q}, we also compare the reconstructed plots of $q(z)$ for our model and the standard $\Lambda$CDM model. The plot demonstrates that our model's evolution of $q(z)$ is consistently compatible with the $\Lambda$CDM model across different datasets. Furthermore, it is evident that the best-fit values of $q_0$ and $z_{tr}$ align well with the predictions of the standard $\Lambda$CDM model. However, as $z$ approaches $-1$, we observe a slight deviation in the evolution of $q(z)$ from the $\Lambda$CDM model, particularly for both $Hz$ and $Hz+BAO$ datasets. The observed behavior of $q(z)$ could potentially be attributed to the choice of the function $B(z)$ in Eq. (\ref{Ez}). As previously mentioned, the term $B(z)$ represents a correction to the $\Lambda$CDM model and plays a significant role in shaping the future evolution of the Universe. 

\begin{figure}[h]
\includegraphics[scale=0.7]{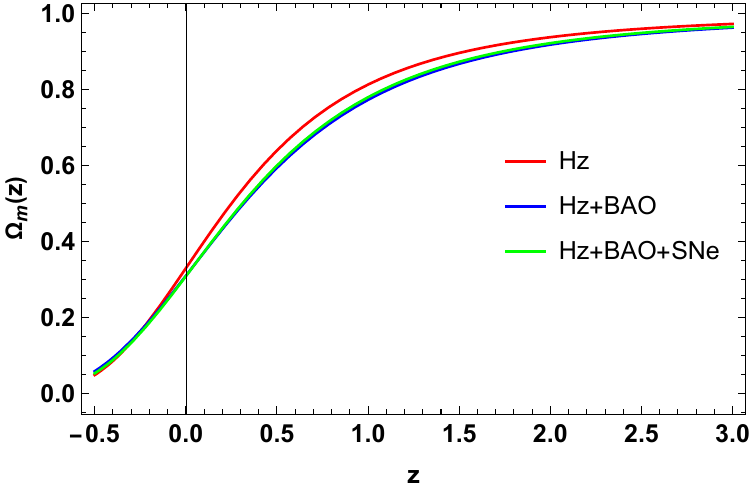}
\caption{This figure shows the evolution of the density parameter for matter $\Omega_{m}(z)$ against $z$ for constrained values from $Hz$, $Hz+BAO$, and $Hz+BAO+SNe$ datasets.}\label{F_Omegam}
\end{figure}

\begin{figure}[h]
\includegraphics[scale=0.7]{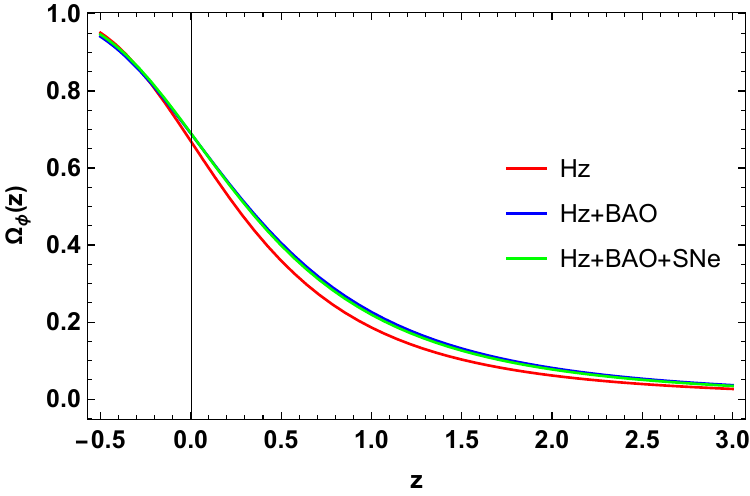}
\caption{This figure shows the evolution of the density parameter for scalar field $\Omega_{\phi}(z)$ against $z$ for constrained values from $Hz$, $Hz+BAO$, and $Hz+BAO+SNe$ datasets.}\label{F_Omegaphi}
\end{figure}

Figs. \ref{F_Omegam} and \ref{F_Omegaphi} illustrate the evolution of the density parameters for matter and the scalar field, respectively. These plots provide important information on the composition of the Universe. Initially, the Universe is predominantly dominated by non-relativistic matter, including dark matter and baryonic matter, while the contribution from the scalar field density parameter remains negligible. As the Universe expands, the density parameter for matter gradually decreases due to the expansion of the Universe. However, with the passage of time, the scalar field's density parameter becomes increasingly significant, eventually surpassing the contribution from matter. This shift in dominance leads to the acceleration of the Universe's expansion, marking a crucial transition in cosmic evolution. In addition, the present-day density parameters are found as $\Omega_{m_0} = 0.33\pm0.11$, $\Omega_{m_0} = 0.310_{-0.038}^{+0.042}$, and $\Omega_{m_0} = 0.309_{-0.038}^{+0.041}$ for the $Hz$, $Hz+BAO$, and $Hz+BAO+SNe$ datasets, respectively. In $\Lambda$CDM model, $\Omega_{m_0} = 0.315\pm0.007$ \cite{Mould,Spergel,Komatsu,Vasey}. It is clearly seen that these values of our model are consistent with those of $\Lambda$CDM.

\begin{figure}[h]
\includegraphics[scale=0.7]{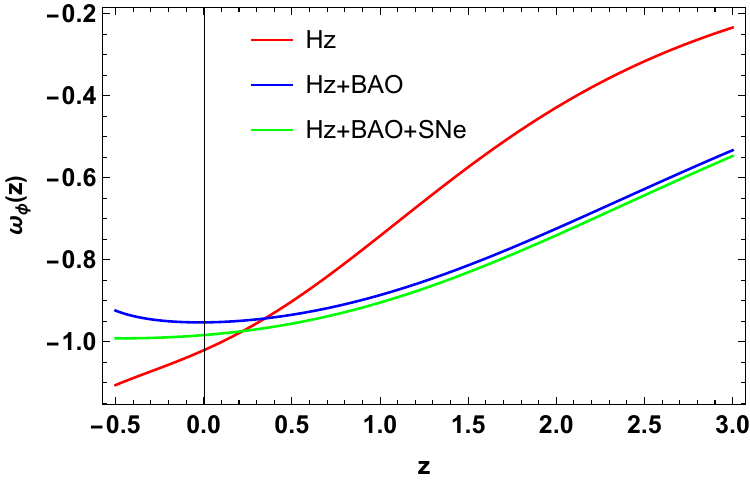}
\caption{This figure shows the evolution of the EoS parameter for scalar field $\omega_{\phi}(z)$ against $z$ for constrained values from $Hz$, $Hz+BAO$, and $Hz+BAO+SNe$ datasets.}\label{F_EoS}
\end{figure}

Fig. \ref{F_EoS} displays the evolution of the EoS parameter of the scalar field $\omega_{\phi}$, which provides crucial insights into the various epochs of accelerated and decelerated expansion of the Universe, as discussed previously. It is clear that the Universe has experienced a transition from quintessence to phantom, and the present EoS of the scalar field lies in the quintessence region ($\omega_{\phi} > -1$) for $Hz+BAO$ and $Hz+BAO+SNe$ datasets, while in the phantom region ($\omega_{\phi} < -1$) for the $Hz$ dataset. At early epochs, the effect of the function $B(z)$ on the EoS parameter is not very significant, but its impact becomes apparent in the future evolution of the Universe. Further, the present values of $\omega_{\phi}(z)$ are found to be $\omega_0=-1.07_{-0.68}^{+0.68}$, $\omega_0=-0.97_{-0.30}^{+0.32}$, and $\omega_0=-0.99_{-0.23}^{+0.23}$ for $Hz$, $Hz+BAO$, and $Hz+BAO+SNe$ datasets, respectively \cite{Garza,Camarena}.

\section{Final Remarks and Perspectives}
\label{sec5}
In the present work, we have conducted an in-depth investigation of the dynamics of the accelerating scenario within the framework of scalar field DE models. Our approach is based on a novel and comprehensive parametrization of the dimensionless Hubble parameter i.e. $E^2(z) = A(z) + \beta (1 + \gamma B(z))$. The functions $A(z)$ and $B(z)$ in our parametrization have been carefully selected to ensure a comprehensive representation of the Hubble parameter over a wide range of redshifts, thereby enhancing the accuracy of our analysis. By employing the chosen parameterization, we derived analytical solutions for various cosmological parameters, including the deceleration parameter, density parameter, and EoS parameter. 

To validate our results, we utilized observational data from diverse sources, such as CC, BAO, and the Pantheon+ datasets. Employing the MCMC methodology, the best-fit values of the model parameters ($H_0,\alpha,\gamma$), along with the corresponding 1-$\sigma$ and 2-$\sigma$ confidence regions, have been determined and are presented in Tab. \ref{tab1} and Fig. \ref{combine}, respectively. The deceleration parameter $q(z)$ was examined in Fig. \ref{F_q} for various redshift values, indicating the transition between decelerated and accelerated phases of the Universe. The analysis of the deceleration parameter reveals that the present values are found to be $q_0=-0.57_{-0.46}^{+0.46}$, $q_0=-0.50_{-0.35}^{+0.36}$, and $q_0=-0.53_{-0.29}^{+0.29}$ for $Hz$, $Hz+BAO$, and $Hz+BAO+SNe$ datasets, respectively. Furthermore, the best-fit values for the transition redshift ($z_{tr}$) are determined as $z_{tr}=0.59_{-0.4}^{+0.4}$, $z_{tr}=0.65_{-0.06}^{+0.08}$, and $z_{tr}=0.65_{-0.08}^{+0.09}$ for $Hz$, $Hz+BAO$, and $Hz+BAO+SNe$ datasets, respectively. In comparing the Hubble constant values obtained for our model with those of the $\Lambda$CDM model, a noteworthy consistency emerges. Our model's Hubble constant values, derived from the $Hz$, $Hz+BAO$, and $Hz+BAO+SNe$ datasets, are remarkably close, with central values of $67.8$, $67.9$, and $68.0$, respectively. On the other hand, the $\Lambda$CDM model, as determined by Planck measurements \cite{Planck2020}, yields a Hubble constant value of $H_0=67.4 \pm 0.5$. The comparison demonstrated the effectiveness of our parametrization in describing the late-time cosmic acceleration and its compatibility with the $\Lambda$CDM model at different redshifts. We reached a similar conclusion using the current values of the density parameters of our model and the $\Lambda$CDM model. We have determined the present-day density parameters as follows: $\Omega_{m_0} = 0.33\pm0.11$, $\Omega_{m_0} = 0.310_{-0.038}^{+0.042}$, and $\Omega_{m_0} = 0.309_{-0.038}^{+0.041}$ for the $Hz$, $Hz+BAO$, and $Hz+BAO+SNe$ datasets, respectively. These values are remarkably close to the corresponding $\Lambda$CDM's density parameter, which is $\Omega_{m_0} = 0.315\pm0.007$. This convergence with the $\Lambda$CDM model indicates a strong agreement with the observed data and underscores the robustness of our findings.

Furthermore, we analyzed the density parameter in Figs. \ref{F_Omegam} and \ref{F_Omegaphi}, which provided valuable information about the relative contributions of different components to the total energy density of the Universe. This enabled us to understand the dominant energy components during different epochs of the Universe's evolution. In addition, the EoS parameter for the scalar field was studied (Fig. \ref{F_EoS}), which is a critical quantity characterizing the nature of DE. Our results shed light on whether the DE behaves as quintessence (with EoS greater than -1) or phantom (with EoS less than -1), indicating whether the Universe's expansion is driven by an evolving scalar field or a cosmological constant, respectively. The present values of EoS are found to be $\omega_0=-1.07_{-0.68}^{+0.68}$, $\omega_0=-0.97_{-0.30}^{+0.32}$, and $\omega_0=-0.99_{-0.23}^{+0.23}$ for $Hz$, $Hz+BAO$, and $Hz+BAO+SNe$ datasets, respectively. 

In conclusion, the model we consider appears to be compatible with $\Lambda$CDM in terms of the evolution and present values of dynamic and kinematic quantities such as the deceleration parameter and density parameters. Therefore, our model is able to explain the expansion evolution of the universe in a manner consistent with observations, without facing the problems that the $\Lambda$CDM model currently faces, and moreover, as we have already mentioned above, it offers the opportunity to analyze in a wider redshift range due to the existence of the function $B(z)$ seen in the parametrization \eqref{Ez}, which can be considered as a correction to the $\Lambda$CDM, and thus providing predictions about future evolution can be considered as the advantages of our model over the $\Lambda$CDM model. Our parametrization of the dimensionless Hubble parameter offers a versatile approach to exploring scalar field DE models \cite{Wang}. The flexibility in choosing the functions $A(z)$ and $B(z)$ allows for addressing various concerns and scenarios within the standard model. In future research, it is valuable to compare our parametrization with alternative models, such as the linear model ($B(z)=z$), the sinusoidal model ($B(z)=\sin(z)$), and the logarithmic model ($B(z)=\log(z+1)$), to assess their respective strengths and weaknesses. These comparative analyses can significantly enrich our understanding of DE and its role in the late-time Universe evolution. Moreover, exploring additional parametrization forms beyond these alternatives offers promising opportunities for achieving a more adaptable approach to modeling cosmic expansion. Furthermore, expanding our analysis to incorporate additional observational datasets and integrating other cosmological probes, such as the CMB radiation data from the Planck mission \cite{CMBdata}, has the potential to yield even more comprehensive and robust constraints on the model parameters. This extended approach would enhance the reliability and depth of our research findings.

\section*{Acknowledgments}
This research is funded by the Science Committee of the Ministry of Science and Higher Education of the Republic of Kazakhstan (Grant No. AP09058240).

\textbf{Data availability} All data used in this study are cited in the references and were obtained from publicly available sources.

\end{document}